\documentclass[dvips,a4paper,11pt]{article}
\title{\textbf{Hyperspherical partial wave theory applied to electron
hydrogen-atom ionization calculation for equal energy sharing
kinematics}}
\date{}
\usepackage{amssymb}
\usepackage{latexsym}
\begin{document}
\setcounter{section}{0}
\renewcommand{\thesection}{\Roman{section}.}
\setcounter{equation}{0}
\renewcommand{\theequation}{\arabic{equation}}
\maketitle
\begin{center}
J. N. Das and S. Paul\\
{\it{Department of Applied Mathematics, University College of
Science, 92, Acharya Prafulla Chandra Road, Calcutta - 700 009,
India}}\\
\end{center}
\begin{center}
K. Chakrabarti\\
{\it{Department of Mathematics, Scottish Church College, 1 \& 3
Urquhart Square,
\\ Calcutta - 700 006, India}}\\
\end{center}
\begin{abstract}
Hyperspherical partial wave theory has been applied here in a new
way in the calculation of the triple differential cross sections
for the ionization of hydrogen atoms by electron impact at low
energies for various equal-energy-sharing kinematic conditions.
The agreement of the cross section results with the recent
absolute measurements of R\"oder \textit {et al} [51] and with the
latest theoretical results of the ECS and CCC calculations [29]
for different kinematic conditions at 17.6 eV is very encouraging.
The other calculated results, for relatively higher energies, are
also generally satisfactory, particularly for large $\Theta_{ab}$
geometries. In view of the present results, together with the fact
that it is capable of describing unequal-energy-sharing kinematics
[35], it may be said that the hyperspherical partial wave theory
is quite appropriate for the description of ionization events of
electron-hydrogen type systems. It is also clear that the present
approach in the implementation of the hyperspherical partial wave
theory is very appropriate.\\\\
\end{abstract}
\noindent
\textbf{PACS Nos 34.80.Dp, 34.50.Fa}\\
e-mail: jndas@cucc.ernet.in\\\\

\pagebreak

\section{INTRODUCTION}
    In the study of electron hydrogen atom ionization collision,
the simplest three-body ionization problem in atomic physics,
there are many attempts for a complete solution but all of these
face tremendous difficulties and have only limited success. Except
for one or two attempts all use time-independent framework. For
accurate information regarding scattering events, one may solve
accurately the Schr\"odinger equation for the scattering states
$\Psi_i^{(+)}$ or $\Psi_f^{(-)}$ [see Newton [1] for their
definition] given by

\begin{equation}
H \Psi_{i,f}^{(\pm)} = E \Psi_{i,f}^{(\pm)}
\end{equation}
taking account of the appropriate boundary conditions.

    Ionization amplitudes may then be obtained either from the flux condition at
infinity or from appropriate projections. In the literature both
$\Psi_i^{(+)}$ and $\Psi_f^{(-)}$ have been widely used. There are
a large number of attempts which strives to solve for
$\Psi_i^{(+)}$.  Among these the most successful attempts are the
various close-coupling calculations [2-4]. In these calculations
$\Psi_i^{(+)}$ are expanded in terms of basis functions and
ionization information are extracted from a solution of the
unknown expansion functions. Another possibility is to expand
$\Psi_i^{(+)}$ in terms of a complete set of functions in the
angular variables. In these regards the attempts of Kato and
Watanabe [5, 6] are remarkable. They used hyperspherical
co-ordinates and expanded $\Psi_i^{(+)}$ in terms of hyper-radius
dependent angular functions. Matching with a wave function, which
satisfies an approximately correct boundary condition, they
obtained with remarkable success, the total ionization cross
sections down to the threshold. However, differential cross
section results of this theory are not known. Very recently
Rescigno and associates made [7, 8] a breakthrough calculation and
reproduced for equal energy-sharing and constant angular
separation $\Theta_{ab}$ of the outgoing electrons, the cross
section results, at low energies, with surprising success. In
these calculations they expanded $\Psi_i^{(+)}$ in terms of
spherical harmonics in four angular variables. Then they converted
the resultant differential equations for the radial functions, in
two radial variables, into a set of difference equations over a
large network in the radial variables-plane. They used a novel
technique. Using a complex scaling procedure they converted the
scattering problem as if into a bound state problem. Then they
solved a huge set (several million) of linear equations using very
special techniques. Ultimately they obtained ionization amplitudes
using the flux condition. Later [9] they confirmed their results
using projection technique. Although the ECS approach reproduced
the equal-energy-sharing, constant-$\Theta_{ab}$ results perfectly
well, results of this approach for unequal-energy-sharing
kinematics are not known. There are also large number of attempts
of using $\Psi_f^{(-)}$ in extracting ionization information. In
such cases projection approach has been generally used. There the
ionization amplitudes are calculated from
\begin{equation}
T_{fi}^s = \langle\Psi_{fs}^{(-)}|V_i|\Phi_i\rangle.
\end{equation}
Brauner, Briggs and Klar [10] and later Berakdar [11] and Berakdar
\textit{et al} [12, 13] made use of this approach. They used
$\Psi_f^{(-)}$ which are asymptotically correct (or nearly so) but
are unlikely to be correct at finite distances. As a consequence
results of these calculations are only moderately accurate.
Moreover there are no systematic tractable way of improving the
results.

    An alternative approach for determining the electron atom
collision cross sections is to solve a coupled set of integral
equations for the off-shell T-matrix elements. Das and associates
[14-16] have used this approach in the study of various electron
hydrogen atom and electron helium atom scattering problems by
solving the resultant equations in a rather crude manner. However,
they always obtained moderately good results. There are also
attempts [17-24] to improve the calculations. Along these lines
the most successful calculations are the convergent close coupling
(CCC) calculations of Bray \textit{et al} [25-30]. In many
contexts they applied the CCC method with surprising success.
Their latest results [29, 30] claim accuracies comparable with the
ECS results at low energies.

    Another promising approach for the electron hydrogen atom
ionization problem is the hyperspherical partial wave approach
[33-34]. Details of this approach are given in [34]. In section II
we also present important features of this approach. Earlier with
an additional approximation of neglecting the coupling effects,
some results were obtained [35, 36] which are qualitatively not
very bad. Recently this approach has been used [37, 38] retaining
fully the coupling effects. In solving the relevant coupled set of
radial wave equations over an initial interval $[0, \Delta]$,
R-matrix [39] approach had been used.  Although the results were
always found to be of the correct magnitude, pseudo-resonance type
behaviour gave much troubles in extracting correct cross section
results. To avoid this problem we use a new approach. This appears
to be very successful and leads to very interesting results both
for equal-energy-sharing constant-$\Theta_{ab}$ kinematics,
equal-energy-sharing asymmetric kinematics, and also for
unequal-energy-sharing kinematics [38]. Thus it appears that
hyperspherical partial wave theory is quite appropriate for the
study of ionization problems of electron-hydrogen type systems.

    Most recently two very broad-based theories have been proposed. One
of these is the time-dependent close coupling theory [40] and the
other is the hyperspherical R-matrix theory [41]. Positions of
these theories are not yet very clear.

\section{ HYPERSPHERICAL PARTIAL WAVE THEORY}

    In the hyperspherical partial wave theory one uses the time-independent
framework. In the time-independent framework the T-matrix element is given by
expression (2) or alternatively by
\begin{equation}
T_{fi} = \langle \Phi_f |V_f|\Psi_i^{(+)}\rangle.
\end{equation}
In these expressions $\Phi_i$ and $\Phi_f$ are the unperturbed
initial and final channel wave functions, satisfying certain exact
boundary condition at infinity and that $V_i$ and $V_f$ are the
corresponding perturbation potentials. For the case of ionization
of hydrogen atoms expression (2) is more appropriate for use,
since in this case asymptotically correct $\Phi_i$ is easily
available. Many use expression (3), including ECS [9] by
projection method, but inappropriately, since the corresponding
$\Phi_f$'s they use do not satisfy the correct boundary condition.
In the hyperspherical partial wave theory $\Psi_f^{(-)}$ is
expanded in terms of hyperspherical harmonics, which are functions
of five angular variables. The corresponding radial waves are
functions of one radial variable, the hyperradius $R$ only. This
proves to be advantageous in numerical computations, since then
the five angular variables range over a bounded compact domain,
while only one variable $R$ ranges over a semi-infinite domain
$[0, \infty)$. It may be noted here that so far nobody could take
account of the exact boundary condition in the asymptotic domain
for the accurate solution of $\Psi_f^{(-)}$. Here we aspire to
take account the exact boundary condition at infinity, in the
limit. This is the most novel feature in the hyperspherical
partial wave theory. Here we may note that two plane waves
$exp(i\vec{p_a} \cdot \vec{r_1})/(2\pi)^{3/2}$ and
$exp(i\vec{p_b}\cdot \vec{r_2})/(2\pi)^{3/2}$ may be decomposed in
partial waves as usual and then these may be combined (using a
formula in Erd\'elyi [42]) to obtain an expansion in terms of
hyperspherical harmonics $\phi_{\lambda}(\omega)$, in five angular
variables $\omega = (\alpha, \theta_1, \phi_1, \theta_2, \phi_2)$.
A symmetrized two-particle plane wave has the expansion [Das, 34]
\begin{eqnarray}
\lefteqn{[exp(i\vec{p_a}\cdot\vec{r_1}+i\vec{p_b}\cdot\vec{r_2})+(-1)^s
exp(i\vec{p_b}\cdot\vec{r_1}+i\vec{p_a}\cdot\vec{r_2})]/(2\pi)^3 } \nonumber\\
& & = 2\sqrt{\frac{2}{\pi}}
\sum_{\lambda}i^{\lambda} \frac{j_{\nu_{\lambda}} (\rho)}{{\rho}^{\frac{3}{2}}}\;
\phi_{\lambda}^{s*}(\omega_0)\;\phi_{\lambda}^{s}(\omega),
\end{eqnarray}
where $\nu_{\lambda}=\lambda +  \frac{3}{2}$ and $\lambda =
l_1+l_2+2n$ ($\lambda$ also denotes the multiplet $(l_1, l_2, n)$
depending on the context). Here $R = \sqrt{r_1^2 + r_2^2}$,
$\alpha = atan(r_2/r_1)$, $\vec{r_1}= (r_1, \theta_1, \phi_1)$,
$\vec{r_2}= (r_2, \theta_2, \phi_2)$. Similarly $P = \sqrt{p_a^2 +
p_b^2}$, $\alpha_0 = atan(p_b/p_a)$, $\vec{p_a}= (p_a, \theta_a,
\phi_a)$, $\vec{p_b}= (p_b, \theta_b, \phi_b)$, and $\rho = PR$,
and $\omega_0 =(\alpha_0, \theta_a, \phi_a, \theta_b, \phi_b)$,

\begin{eqnarray}
\phi_{\lambda}^{s}(\omega) & = & \frac{1}{\sqrt{2}}\{
P_{l_1l_2}^n(\alpha) {\mathcal{Y}}_{l_1l_2}^{LM}(\hat{r_1},
\hat{r_2}) + (-1)^{l_1+l_2-L+S+n}
P_{l_2l_1}^n(\alpha){\mathcal{Y}}_{l_2l_1}^{LM}(\hat{r_1},
\hat{r_2})\},
l_1 \ne l_2 \nonumber \\
& = & \frac{1}{2}\{1 + (-1)^{-L+S+n}\}P_{ll}^n(\alpha)
{\mathcal{Y}}_{ll}^{LM}(\hat{r_1}, \hat{r_2})\}, \; \mathrm{for}
\; l_1 = l_2 = l,
\end{eqnarray}
and a corresponding expression for $\phi_{\lambda}^{s}(\omega_0)$
(similar expressions may be easily derived for product of more
than two plane waves).

    Now the symmetrized wave $\Psi_{fs}^{(-)}$ may be expanded in
terms of symmetrized hyperspherical harmonics
$\phi_{\lambda}^{s}$'s as
\begin{equation}
\Psi_{fs}^{(-)}(R, \omega)=2\sqrt{\frac{2}{\pi}} \sum_{\lambda}
\frac{F_{\lambda}^s(\rho)}{{\rho}^{\frac{5}{2}}}\;
\phi_{\lambda}^s(\omega),
\end{equation}
where $F_{\lambda}^{(s)}$ satisfy an infinite coupled set of equations
\begin{equation}
\Big[ \frac{d^2}{d R^2} + P^2 -
\frac{\nu_{\lambda}\,(\nu_{\lambda}+1)} {R^2}\Big]
F_{\lambda}^{(s)}(R)+ \sum_{\lambda'} \frac{2P\; \alpha_{\lambda
\lambda'}^s}{R} \, F_{\lambda'}^{(s)}(R) = 0,
\end{equation}
for each symmetry s ($s = 0$ for singlet and $s = 1$ for triplet)and for each
total angular momentum $L$ (and its projection $M$, and so also for a definite
parity $\pi$). In the above expression
\begin{displaymath}
\alpha_{\lambda \lambda'}^s = -\langle \phi_{\lambda}^s|C| \phi_{\lambda'}^s
\rangle/P, \; \mathrm{and}
\end{displaymath}
\begin{displaymath}
C=- \frac{1}{cos\alpha}-\frac{1}{sin\alpha}+\frac{1}{|\hat{r_1}cos\alpha
-\hat{r_2}sin\alpha|}. \nonumber
\end{displaymath}
The equations (7) have to be solved over the semi-infinite domain
$[0, \infty)$. Solution in the asymptotic domain is simple. The
equations have nice asymptotic solutions. One may note that
because of conservation rules the equations (7) are decoupled into
sets for fixed $\mu = (L, S, \pi)$ and different $N = (l_1, l_2,
n)$. So we set $F_{\lambda}^{(s)} \equiv f_N^{\mu}$ and, for the
set with fixed $\mu$ (and dropping $\mu$ from $f_{N}^{\mu}$) we
can write equations (7) as
\begin{equation}
\Big[ \frac{d^2}{dR^2} + P^2 - \frac{\nu_N\,(\nu_N+1)}
{R^2}\Big]f_N + \sum_{N'} \frac{2P\;\alpha_{NN'}}{R} \,
f_{N'} = 0,
\end{equation}
for a solution over the finite domain and
\begin{equation}
\Big[ \frac{d^2}{d\rho^2} +1 - \frac{\nu_N\,(\nu_N+1)}
{\rho^2}\Big]f_N + \sum_{N'} \frac{2\;\alpha_{NN'}^s}{\rho} \,
f_{N'} = 0,
\end{equation}
for solution over an asymptotic domain, say $[R_0,\infty)$. Next
we consider the solution problem first over an asymptotic domain
$[R_0,\infty)$  and then over the finite domain $[0, R_0]$.\\\\

\noindent
\textbf{A. Solution in an asymptotic domain}\\

    The equations (9) have two sets of solutions [34] in an asymptotic
domain $[R_0, \infty)$ of the form given by
\begin{equation}
f_{snN}^{(k)}(\rho) = \sum_{\ell=0}^{\infty}\; \frac{a_{kN}^{(\ell)}sin\;
\theta_k}{\rho^\ell} + \sum_{\ell=0}^{\infty}\; \frac{b_{kN}^{(\ell)}cos\;
\theta_k}{\rho^\ell},
\end{equation}
\begin{equation}
f_{csN}^{(k)}(\rho) = \sum_{\ell=0}^{\infty}\;
\frac{c_{kN}^{(\ell)}sin\; \theta_k}{\rho^\ell} +
\sum_{\ell=0}^{\infty}\; \frac{d_{kN}^{(\ell)}cos\;
\theta_k}{\rho^\ell},
\end{equation}
where $\theta_k = \rho + \alpha_k\, ln\,2\rho$ and $\alpha_k$ is the k-th
eigen value of the charge matrix $A = (\alpha_{NN'})$ and that the coefficients
$a_{kN}^{(l)},\; b_{kN}^{(l)}, \; c_{kN}^{(l)}\; \mathrm{and}\; d_{kN}^{(l)}$ are determined
from recurrence relations. Thus the coefficients $a_{kN}^{(l)}$ and
$b_{kN}^{(l)}$ are determined from the relations -
\begin{eqnarray}
2[(A_k)^2 + l^2 I]\mathbf{X}_k^{(l)} &=& [A_k \Lambda_k -
l(l-1)A_k -l(2l-1) \alpha_k I]\mathbf{X}_k^{(l-1)} \nonumber\\
& &- [(2l-1)\alpha_k A_k + l\Lambda_k -
\l^2(l-1)I]\mathbf{Y}_k^{(l-1)}
\end{eqnarray}
and
\begin{eqnarray}
2[(A_k)^2 + l^2 I]\mathbf{Y}_k^{(l)} &=& [A_k \Lambda_k -
l(l-1)A_k -l(2l-1) \alpha_k I]\mathbf{Y}_k^{(l-1)} \nonumber\\
& &+ [(2l-1)\alpha_k A_k + l
\Lambda_k-\l^2(l-1)I]\mathbf{X}_k^{(l-1)}
\end{eqnarray}
where the N-th components of vectors $\mathbf{X}_k^{(l)}$ and
$\mathbf{Y}_k^{(l)}$ are given by
$$(\mathbf{X}_k^{(l)})_N = a_{kN}^{(l)}, \;(\mathbf{Y}_k^{(l)})_N =
b_{kN}^{(l)}$$
and where
$$ A_k = A - \alpha_k I, \; (\Lambda_k)_{NN'}
= [{\alpha_k}^2 + \nu_N(\nu_N + 1)] \delta_{NN'}.$$ The initial
vectors $\mathbf{X}_k^{(0)}$ and $\mathbf{Y}_k^{(0)}$ are given by
$$\mathbf{X}_k^{(0)} = \mathbf{X}_k, \; \mathbf{Y}_k^{(0)} = 0,$$
$\mathbf{X}_k$ being the k-th eigen vector of the charge matrix
$A$ and  $I$ is the identity matrix. Solution for $c_{kN}^{(l)}$'s
and $d_{kN}^{(l)}$'s are similarly obtained from the above
recurrence relations after setting $\mathbf{X}_k^{(0)} =0$ and
$\mathbf{Y}_k^{(0)} = \mathbf{X}_k$. In this way we get solution
vectors $\bf{f_{snN}^{(k)}}$ and $\bf{f_{csN}^{(k)}}$ of equations
(10) and (11).\\

\textbf{B. Solution over a finite domain}\\

    Here we consider the solution of equations (8) over a finite domain
$[0, R_0]$. Away from the origin, solution of the equations is
easy. A Taylors series expansion method suffices for getting
arbitrarily accurate solutions. The main difficulty is in starting
the solution from the origin. Near origin the equations have
analytic solutions ( Fock [43] )but these are too complicated to
be useful in numerical computations. In our earlier calculations
[37, 38] we used R-matrix approach for getting solutions over an
initial interval $[0, \Delta]$  (with $\Delta$ suitably chosen).
But this approach faces difficulties as pseudo-resonance type
behavior appears giving much troubles in determining appropriate
solutions. To avoid such difficulties we consider here a new
approach. For the interval $[0, \Delta]$ we consider a boundary
value problem . The solution vector $\mathbf{f_0^{(k)}(R)}$ is
assumed to have a value $0$ at the origin and the k-th column of
the $N_{mx} \times N_{mx}$ identity matrix at $R = \Delta$. We
divide the interval $[0, \Delta]$ into $m$ subintervals and use a
five-point difference formula
\begin{eqnarray}
f_N^{''}(R_k)=\frac{1}{12\,h^2}[-f_N(R_{k-2})+16f_N(R_{k-1})-30f_N(R_k)
& + & 16f_N(R_{k+1}) - f_N(R_{k+2})]  \nonumber\\
& & + \{\frac{h^4}{90}f_N^{(vi)}(\xi)\}
\end{eqnarray}
for $k = 2, 3, \cdots , m-2$, and a formula
\begin{eqnarray}
f_N^{''}(R+h') &=& 2f_N^{''}(R+2h') - f_N^{''}(R+3h') + \frac{1}{h^2} [f_N(R)
- 4f_N(R+h') + 6f_N(R+2h') \nonumber\\
& & -  4f_N(R + 3h') + f_N(R + 4h') ] +
\{-\frac{h^4}{12}f^{(vi)}(\xi)\}.
\end{eqnarray}
with $R = R_0$, $h^{'} = h$ for the equation at $R = R_1$ and $R =
R_m$ , $h^{'} = -h$ for the equation at $R = R_{m-1}$.For
continuing solutions from $\Delta$ onward we need first order
derivatives at $\Delta$. For this we use the simple formula
\begin{eqnarray}
f_N^{'}(R_m) &=& [-f_N(R_{m-4}) + 24f_N(R_{m-2}) - 128f_N(R_{m-1})
+ 105f_N(R_m)]/(84h) \nonumber\\ & & + \frac{2h}{7} f_N^{''} (R_m)
 + \{-\frac{4h^4}{105}f_N^{(v)}(\xi)\}.
\end{eqnarray}
The resultant linear set of equations ultimately take the form
\begin{equation}
S\mathbf{Y^{(k)}} = \mathbf{b},
\end{equation}
where $S$ is a sparse matrix. We solve these equations by the
biconjugate gradient method [44] using routines given in [45].
With a suitable choice of a preconditioner the iterations smoothly
converge (with a few hundred iterations or even less) to five or
six decimal places for a suitable choice of error limit (say, 1 in
$10^7$ parts). In this way $N_{mx}$ solution vectors
$\mathbf{f_0^{(k)}}$ are determined over $[0, \Delta]$. The
solutions are next continued over $[\Delta, R_0]$ by Taylors
expansion method with stabilization [47] after suitable steps,
giving solution vectors
$\mathbf{f_0^{(k)}}$ over $[0, R_0]$.\\

\textbf{C. Matching of the solutions: Determination of $\Psi_f^{(-)}$}\\

    For finding the physical solution vectors $\mathbf{f_{ph}}$ and the
scattering state $\Psi_{fs}^{(-)}$ we first define solution
matrices $\mathrm{f_0}$, $\mathrm{f_{sn}}$ and $\mathrm{f_{cs}}$,
by putting side by side the corresponding solution vectors
$\mathbf{f_0^{(k)}}$, $\mathbf{f_{sn}^{(k)}}$,
$\mathbf{f_{cs}^{(k)}}$, for $k = 1, 2, \cdots ,N_{mx}$. Then the
physical solution vector $\mathbf{f_{ph}}$ may be defined over
$[0, R_0]$ by

\begin{equation}
\mathbf{f_{ph}}(R) = {\sum_{k=1}}^{N_{mx}} g_k \mathbf{f_{0}^{(k)}}(R)
\end{equation}

and over $[R_0, \infty)$ by\\

\begin{equation}
\mathbf{f_{ph}}(R) = {\sum_{k=1}}^{N_{mx}} c_k \mathbf{f_{sn}^{(k)}}(\rho) +
{\sum_{k=1}}^{N_{mx}} d_k \mathbf{f_{cs}^{(k)}}(\rho).
\end{equation}
$2N_{mx}$ of the $3N_{mx}$ unknown coefficients are now determined by matching
values (and first order derivatives) of the two sets of solutions at a point
$R_0$ where all the solutions are valid. The remaining $N_{mx}$ unknown
coefficients are then determined from the demand that  $\Psi_{fs}^{(-)}$
actually satisfies the appropriate boundary condition. To facilitate the
computations we first define the K-matrix through the relation
\begin{equation}
f_0 \cdot B = f_{sn} +f_{cs} \cdot K
\end{equation}
where B is some unknown constant matrix. (The K-matrix thus
defined is a little different from the one usually defined. But in
any case it should be symmetric.) The K-matrix is then determined
by matching values and first order derivatives of the two sides of
equation (20). Then in the asymptotic domain one has
\begin{eqnarray}
\mathbf{f_{ph}} & = & (f_{sn} + f_{cs} \cdot K) \cdot
\mathbf{c}  \nonumber\\
       & = & f_{sn}\cdot \mathbf{c} + f_{cs} \cdot \mathbf{d},
\end{eqnarray}
where
\begin{equation}
\mathbf{d} = K \cdot \mathbf{c}.
\end{equation}
Finally $\mathbf{f_{ph}}$ is completely determined once the vector
\textbf{c} is determined. Now \textbf{c} is determined from the
consideration that $\Psi_{fs}^{(-)}$ is asymptotically a distorted
plane wave (representing the two outgoing electrons) plus incoming
waves only. So we equate coefficients of the outgoing wave
$exp(i\rho)$ of both $\Psi_{fs}^{(-)}$ and the symmetrized plane
wave (4) (except for the distorting terms $exp(i\alpha_k ln
2\rho)$. This gives
\begin{equation}
\mathbf{c} = \Big[I + iK \Big]^{-1} \, \mathbf{P}
\end{equation}
where
\begin{displaymath}
\mathbf{P} = -e^{\frac{i\pi}{4}}\;X^{-1}\; \mathbf{\Phi}^{(s)*}(\omega_0),
\end{displaymath}
where $X$ is the matrix consisting of eigen vectors of the charge
matrix $A$ (and is non-singular) and
$\mathbf{\Phi}^{(s)*}(\omega_0)$ is given by
\begin{equation}
\mathbf{ {\Phi}^{s*}(\omega_0)}=
\left(\begin{array}{c}
{\phi^{s*}}_1(\omega_0)\\
\; \vdots\\ {\phi^{s*}}_{N_{mx}}(\omega_0)
\end{array} \right).
\end{equation}
In this way the physical radial vectors ${\bf{f_{ph}}}(R)$ are determined
for each $\mu = (L, S, \pi)$ and ultimately the full (but approximate)
scattering state $\Psi_{fs}^{(-)}$ is obtained.

    Substituting this expression in equation (2) one obtains the
scattering amplitude in the form
\begin{equation}
f^{s}(\omega_0)=\frac{1}{2\pi} T_{fi}^{s} = \frac{1}{2\pi}
{\sum_N}^{N_{mx}} C^s(N)\phi_N^s(\omega_0)
\end{equation}
The triple differential cross section is then given by
\begin{equation}
\frac{d^3\sigma}{dE_a d\Omega_a d\Omega_b}=\frac{(2\pi)^4p_ap_b}{p_i}
\Big\{ \frac{1}{4}|T_{fi}^{(0)}|^2 + \frac{3}{4}|T_{fi}^{(1)}|^2 \Big\}.
\end{equation}
By increasing the number of channels $N_{mx}$ for each $\mu = (L, S, \pi)$
one may expect to obtain converged cross section results.

\section{PRESENT CALCULATION}

    In our present calculation there are two important parameters
$\Delta$ and $R_0$ which are needed to be judiciously chosen. The
parameter $\Delta$ of the initial interval [0, $\Delta$], for a
solution of the radial equations (8), has been chosen to be 5 a.u.
for all the energies considered here. The results do not depend on
$\Delta$ for small variations (of a few a.u.) about this value. On
the other hand the choice of the parameter $R_0$, the asymptotic
range parameter, is very crucial. Without its appropriate choice
the asymptotic series solutions (10) and (11) are unlikely to
converge. Here it is found that for convergent asymptotic series
solutions $R_0$ is needed to be such that
$R_0\sim\frac{1}{\sqrt{E}}$, where E is the energy in the final
channel. Thus for energies of 30 eV, 25 eV, 19.6 eV, 17.6 eV, and
15.6 eV this range parameter $R_0$ may be chosen greater than the
values 60 a.u., 70 a.u., 90 a.u., 120 a.u. and 150 a.u.
respectively. We have chosen $R_0$ around these values in our
calculations. However for the computation of single differential
cross section (SDCS) it is necessary for converged results to vary
$R_0$, and extrapolate, as in ECS calculation [8] for
$R_0\rightarrow\infty$. Our limited computational resources
restrict us to take single $R_0$ value for each energy.Moreover
for arbitrary large $R_0$ unwanted errors are likely to make the
results erroneous. So some optimum choice of $R_0$ has to be made
for each energy with a few trials. In our present computations
this has been done.
 Next we consider the choice of L values for inclusion in the
calculations for different energies. For 15.6 eV energy, values of
L upto 5 proved sufficient. For 17.6 eV calculations values of L
upto 7 are found to be necessary. For the other energies
considered here, values of L upto 9 have been included. For fixed
($L,S,\pi$) the number of channels, the number of states with
different ($l_1,l_2,n$) triplets, which have been included, were
chosen suitably for fairly converged results. In any case for
fixed ($l_1,l_2$) pairs n was varied from 0 to 9. In this way
convergence with respect to n is obtained. The number of
($l_1,l_2$) pairs, which have been included, are somewhat less for
$L \ge 2$ compared to those in the ECS calculation . These pairs
are chosen more or less in the order as in ECS calculations (a
little different from those of hyperspherical calculations of Kato
and Watanabe [6]). However there could be some better choice. For
lower energies convergence with respect to the increase in channel
size is rather smooth. It is not so for relatively higher energies
of 25 eV or 30 eV. Nevertheless we have obtained nearly converged
results in the cases considered. All the results presented here
are more or less based on 50 channels calculations. Most of the
calculations, reported here, were done on Pentium -III PCs.
Calculations for 15.6 eV energy could not be done on PCs. Thus the
results for 15.6 eV and 17.6 eV, which are presented here, have
been derived from calculations on a SUN server. It may be further
added here that for 15.6 eV the SDCS results show that for equal
energy sharing case the calculated SDCS value is about twice the
expected value, although the calculated total cross section
appears correct (see table I). This is unacceptable. In any case
the various cross section results for this energy have been
multiplied by a factor 0.5 before presentation in the figures. For
other energies, however, we have nearly the correct SDCS values
for equal energy sharing situations. Calculation on a larger scale
with larger values of $R_0$, and with more precise solution of the
equations (17), may decide absolutely the normalization question
of the measured results of R\"oder \textit{et al} [48] for 15.6 eV
energy. Cross section results for 15.6 eV energy have been
included here for the sake of completeness.

\section{RESULTS}

\textbf{A. Triple Differential Cross Section for Constant
$\Theta_{ab}$Geometry}\\

    The triple differential cross section
results for equal-energy-sharing constant $\Theta_{ab}$ geometries
are presented in figures 1(a) for 15.6 eV energy, in figure 2(a)
for 17.6 eV energy and in figures 3, 4 and 5 for energies of 19.6
eV, 25 eV and 30 eV. In these figures we have presented the
theoretical results of CCC calculations [26, 28, 29, 30, 51] and
of ECS calculations [8, 50]. Here we have also included the
absolute measured values of R\"oder \textit{et al} [47, 48] for
15.6 eV and the most recent re-measured (with necessary
inter-normalization) values of R\"oder \textit{et al} [51] for
17.6 eV energy. For other energies the measured results [47] are
only relative and are normalized as in [9]. Our results are
generally comparable with the ECS results in magnitude. For 17.6
eV our present results appear most interesting. These are even
somewhat better compared to the ECS and CCC results for
$\Theta_{ab}$ = $150^0$ and $180^0$. For these values of
$\Theta_{ab}$, the 15.6 eV results also appear good , particularly
in shapes, but confirmation by larger scale calculation is
necessary. The 19.6 eV results also appear to be very good. For
other energies our results appear less satisfactory in comparison
with the ECS and CCC results.\\

 \textbf{B. Triple Differential Cross Sections for Fixed
 $\theta_a$  Geometry}\\

    In figures 1(b) and 2(b) we have compared our results for
equal-energy-sharing asymmetric geometries with absolute measured
values of R\"oder \textit {et al} [48] for 15.6 eV and R\"oder
\textit {et al} [51] for 17.6 eV, in which one of the outgoing
electrons is observed in a fixed direction while that of the other
one is varied. In these cases we again compare our results with
the calculated results of ECS and CCC theories. Here our results
also appear to be quite good, particularly for 17.6 eV in view of
the most recent measurements. For $\theta_a$ = $-30^o$ at 15.6 eV,
the peak position of our calculated curves are little shifted to
the right. Otherwise all the results of the present calculation
appear satisfactory.\\

\textbf{C. Triple Differential Cross Sections for Symmetric
Geometry}\\
     In figures 1(c) and 2(c) we have presented TDCS
results for symmetric appearance of the two outgoing electrons
relative to the incident electron direction, for 15.6 eV and 17.6
eV incident electron energies, for which there are again absolute
measured results [48, 51]. For 15.6 eV energy our results agree
qualitatively with the experimental results. Here a 70-channel
calculation has been found to be necessary. For 17.6 eV our
results do not appear very good. For 15.6 eV energy both the ECS
and CCC theories underestimate the cross section results
considerably. For 17.6 eV energy, however, both  ECS and CCC
theories give good overall representation.\\

 \textbf{D. Integrated Cross Sections and the Spin Asymmetry Parameter }\\
    The parabolic fitted curves to our computed single differential
cross sections data are generally close to the ECS extrapolated
curves but our raw data which could be calculated, as in ECS flux
method, away from the two ends of the energy intervals widely
differ from ECS or CCC (wherever available) curves. However, the
computed total integrated cross sections, with suitable
extrapolation from these are generally good. Here in table I we
have presented values of integrated cross sections $\sigma_I =
(\sigma_s + 3\sigma_t)$/4 and the spin asymmetry parameter A =
($\sigma_s - \sigma_t$)/($\sigma_s + 3\sigma_t$) where $\sigma_s$
and $\sigma_t$ are the singlet and the triplet cross sections,
together with values of ECS theory by flux approach [8] and those
of CCC theory and the experimental values . The integrated cross
sections agree with the experimentally measured values of Shah
\textit {et al} [52] within about $20\%$ . The spin asymmetry
parameter A agrees, however, excellently with the measurements
[53, 54].\\

    Next we note down the shortcomings and difficulties associated
with the present approach. The first point to note is that it may
not be possible in this approach to get reliable cross section
results for extreme asymmetry, as in ECS flux approach, for one of
the outgoing electrons sharing very small energy values compared
to the other. 'Contamination with high Rydberg states', as in ECS
calculation [8] gives wrong results for finite values of $R_0$ in
such cases. Extrapolation to $R_0 = \infty$ may only lead to
reliable results in those cases. This may require larger
computational resources. Another difficulty to be noted is the
appearance of a few large eigen-values of the charge matrix for
large-channel calculations. In such cases computational strategies
are needed to be reviewed. In our calculations this has occurred
in a few cases. In such cases we simply cut-short in magnitude
these one or two large eigen values to the neighboring ones.
However a better approach may be necessary to tackle such
problems. No other difficulties appear worth mentioning. For a
fully converged results inclusion of more channels (about 100 or a
little more) may be required with appropriate choice of
($l_1,l_2$) pairs (say, as in ECS calculation) and with further
stabilization. However these are subjects of further studies
requiring more computational resources and time.

\section{CONCLUSIONS}

    The results of the present calculation fairly display the
capability of the hyperspherical partial wave theory in
representing results for equal-energy-sharing kinematical
conditions at low energies. The new approach that has been used in
the implementation of the hyperspherical partial wave theory
appears very appropriate. The computed cross section results are
observed to be very satisfactory. If one recalls the capability of
the theory to describe the ionization collisions for
unequal-energy-sharing asymmetric kinematic conditions (as
indicated in [38]) also then the capability of the hyperspherical
partial wave theory towards a complete description of the electron
- hydrogen - atom ionization problem is well demonstrated.
Considering the computational facilities used (Pentium - III PCs
and a SUN Enterpriser 450 server) success of the present
calculation is appreciable. For fully converged results better
computational facilities may be required.
    The theory may easily be applied in the study of ionization of
hydrogen-like ions with a little change in the definition of the
wave function $\Phi_{i}$ and the interaction potential $V_i$. The
theory may also be extended for application to the
double-ionization of helium atom or helium-like ions or to other
multi electron ionization processes.

\section{ACKNOWLEDGMENTS}

    The authors are grateful to Igor Bray for providing them with
the CCC results, the experimental results of R\"oder \textit{et
al} and preprints of relevant papers in electronic form. They are
grateful to T. N. Rescigno and M. Baertschy for sending  the ECS
results electronically. Special thanks are also due to M.
Baertschy for providing Matlab scripts which helped in drawing the
figures. S. Paul is grateful to CSIR for providing a research
fellowship.\\

\pagebreak

\noindent

\underline{\bf{References}}
\begin{flushleft}

[1] R. G. Newton, \textit{Scattering Theory of Waves and Particles, McGraw-Hill}, NY (1966).\\

[2] E. P. Curran and H. R. J. Walters, J. Phys. B\textbf{20}, 337 (1987); see also
    E. P. Curran, C. T. Whelan and H. R. J. Walters, J. Phys. B\textbf{24}, L19(1991).\\

[3] K. Bartchat, E. T. Hudson, M. P. Scott, P. G. Burke, and V. M. Burke, J. Phys. B
    \textbf{29}, 115 (1996).\\

[4] K. Bartschat and I. Bray, J. Phys. B\textbf{29}, L577 (1996).\\

[5] D. Kato and S. Watanabe, Phys. Rev. Lett \textbf{74}, 2443 (1995).\\

[6] D.Kato and S Watanabe, Phys. Rev.A \textbf{56}, 3687 (1997).\\

[7] T. N. Rescigno, M. Baertschy, W. A. Isaacs, and C. W. McCurdy, Science \textbf{286},
    2474 (1999).\\

[8] M. Baertschy, T. N. Rescigno, W. A. Isaacs, X. Li, and C. W. McCurdy, Phys. Rev.
    A \textbf{63}, 022712 (2001).\\

[9] M. Baertschy, T. N. Rescigno, and C. W. McCurdy, Phys. Rev. A\textbf{64}, 022709 (2001).\\

[10] M. Brauner, J. S. Briggs, and H. Klar, J. Phys. B \textbf{22}, 2265 (1989).\\

[11] J. Berakdar, Phys. Rev. A \textbf{53}, 2314(1996).\\

[12] J. Berakdar, J. R\"oder, J. S. Briggs, and H. Ehrhardt, J. Phys. B\textbf{29}, 6203(1996).\\

[13] J. Berakdar, J. S. Briggs, I. Bray, and D. V. Fursa, J. Phys. B\textbf{32}, 895 (1999).\\

[14] J. N. Das, J. Phys. B \textbf{11}, L195 (1978).\\

[15] J. N. Das, Phys. Lett A \textbf{69}, 405 (1979); ibid \textbf{83}, 428 (1981).\\

[16] J. N. Das, R.K.Bera, and B. Patra, Phys. Rev. A \textbf{23}, 732 (1981).\\

[17] J. N. Das and A. K. Biswas, Phys. Lett A \textbf{78}, 319 (1980).\\

[18] J. N. Das and A. K. Biswas, J. Phys. B \textbf{14} ,1363 (1981).\\

[19] J. N. Das and P. K. Bhattacharyya, Phys. Rev. A \textbf{27}, 2876 (1983).\\

[20] J. N. Das and N. Saha,  J. Phys. B\textbf{14}, 2657 (1981).\\

[21] J. N. Das and N. Saha, Pramana- J. Phys. \textbf{18}, 397 (1982).\\

[22] J. N. Das, A. K. Biswas, and N. Saha, Aust. J. Phys. \textbf{35}, 393 (1982).\\

[23] J. N. Das and A. K. Biswas, Czeck. J. Phys. B \textbf{38}, 1140 (1988).\\

[24] S. P. Khare and Kusum Lata, Phys. Rev. A \textbf{29}, 3137
(1984); S. P. Khare and Satya Prakash, Phys. Rev. A \textbf{32},
2689 (1985).\\

[25] I. Bray, D. A. Konovalov, I. E. McCarthy, and A. T.
Stelbovics, Phys. Rev. A\textbf{50}, R2818 (1994).\\

[26] I. Bray, J. Phys. B \textbf{32}, L119(1999) ; J. Phys. B \textbf{33}, 581 (2000).\\

[27] I Bray, Aust. J. Phys. \textbf{53}, 355 (2000).\\

[28] I Bray,  D. V. Fursa, A. S. Kheifets, and A. T. Stelbovics
J. Phys. B\textbf{35}, R117 (2002).\\

[29] J. R\"oder, M. Baertschy, and I Bray, Phys. Rev. A (2002) (to be published).\\

[30] I Bray, Phys. Rev. Lett. (2002) (to be published).\\

[31] G. Bencze and C. Chandler, Phys. Rev. A \textbf{59}, 3129 (1999).\\

[32] S. Jones and D. H. Madison, Phys. Rev. A \textbf{63}, 042701 (2000).\\

[33] J. N. Das, Aust. J. Phys. \textbf{47}, 743 (1994).\\

[34] J. N. Das, Pramana- J. Phys. \textbf{50}, 53 (1998).\\

[35] J. N. Das and K. Chakrabarti, Pramana- J. Phys. \textbf{47}, 249 (1996).\\

[36] J. N. Das and K. Chakrabarti, Phys. Rev. A \textbf{56}, 365 (1997).\\

[37] J. N. Das, Phys. Rev. A \textbf{64}, 054703 (2001).\\

[38] J. N. Das, J. Phys. B  \textbf{35}, 1165(2002).\\

[39] P. G. Burke and W. D. Robb, Adv. At. Mol. Phys. \textbf{11}, 143 (1975).\\

[40] M. S. Pindzola and F Robicheaux, Phys. Rev. A \textbf{55},
4617 (1997); ibid \textbf{57}, 318 (1998); ibid \textbf{61}, 052707 (2000).\\

[41] L. Malegat, P. Selles, and A. Kazansky, Phys. Rev. A \textbf{60}, 3667 (1999).\\

[42] A. Erdelyi, \textit{Higher Transcendental Functions},
Vol.II, p.99. McGraw-Hill, NY (1953).\\

[43] V. Fock, K. Norske Vidensk. Selsksk. Forh \textbf{31}, 138 (1958).\\

[44] R. Fletcher, \textit{Numerical Analysis Dundee, 1975, Lecture
Notes in Mathematics},
Vol. 506, Eds. A. Dold and B. Eckmann, Springer- Verlag, Berlin, pp 73-89 (1976).\\

[45] W. H. Press, S. A. Teukolsky, W. T. Vellering, and B. P.
Flannery, \textit{Numerical
Recipes in Fortran}, p77, 2nd ed., Cambridge University Press (1992).\\

[46] B. H. Choi and K. T. Tang, J. Chem. Phys. \textbf{63}, 1775 (1975).\\

[47] J. R\"oder, J. Rasch. K. Jung, C. T. Whelan, H. Ehrhardt, R.
J. Allan, H. R. J. Walters, Phys. Rev. A \textbf{53}, 225 (1996).\\

[48] J. R\"oder, H. Ehrhardt, C. Pan, A. F. Starace, I. Bray, and
D. Fursa, Phys. Rev. Lett. \textbf{79}, 1666 (1997).\\

[49] I. Bray, private communications.\\

[50] M. Baertschy and T. N. Rescigno, private communications.\\

[51] J. R\"oder, M. Baertschy and I. Bray, Phys. Rev A {\bf{45}},
2951(2002).\\

[52] M. B. Shah, D. S. Elliot, and H. B. Gilbody, J. Phys. B
{\bf{20}}, 3501(1987).\\

[53] D. M. Crowe, X. Q. Guo, M. S. Lubell, J. Slevin, and M.
Eminyan, J. Phys. B {\bf{23}}, L325(1987).\\

[54] G. D. Fletcher, M. J. Alguard, T. J. Gray, P. F. Wainwright,
M. S. Lubell, W. Raith and V. W. Hughes, Phys. Rev. A {\bf{31}},
2854(1985).
\end{flushleft}

\pagebreak

\noindent

\underline{\bf{TABLE}}\\

 Table I. Total integrated ionization cross sections
(a.u.) and the spin asymmetry parameter. The experimental values
of cross sections are those of Shah \textit{et al} [52] (the
starred numbers are from extrapolation). ECS results are from [8]
and the CCC results are from [4]. In the experimental results of
the asymmetry parameter of Crowe {\it{et al}} [53]  and Fletcher
{\it{et al}} [54] presented here, the numbers with superscript +
or $-$ denote the available result just a little above or below
the energy considered (for the exact energy values the
corresponding references are to be seen). For 15.6 eV energy,ECS
results of earlier calculation [8] are not available. So for this
energy we present results from [9] and indicate it so in the table.\\
\begin{tabular}{lcccccc}
 \hline
 \hline
  E$_i$(eV) & 30 & 25 & 19.6 & 17.6 & 15.6 \\
 \hline
  \textbf{Total Integrated Cross Sections}\\
  \hline
  Present: & 2.13 & 1.82 & 1.14 & 0.83 & 0.49 \\
  ECS:  & 1.79 & 1.62 & 1.09 & 0.80 & 0.36 [9] \\
  CCC:  & 1.92 & 1.57 & 1.01 & 0.75 & 0.38\\
  Expt.: & 1.81$^*$ & 1.55$^*$ & 1.00 & 0.74 & 0.39 \\
  \hline
  \textbf{Spin Asymmetry}\\
  \hline
  Present: & 0.31 & 0.41 & 0.47 & 0.55 & 0.48 \\
   ECS: & 0.42 & 0.45 & 0.51 & 0.51 & 0.52 [9]\\
   CCC: & 0.41 & 0.43 & 0.50 & 0.51 & 0.53\\
   Expt.:\\
   Crowe & 0.28 & 0.39$^-$ & 0.42$^-$ & 0.47$^-$ & 0.50$^-$\\
   Fletcher & 0.31 & 0.41$^-$ & 0.40$^+$ & 0.50$^-$ &
         0.48$^-$\\
  \hline
\end{tabular}\\
\pagebreak

\underline{\bf{Figure Captions}}\\

\noindent
    \textbf{Figure 1(a).} TDCS in coplanar equal-energy-sharing constant
angular separation $\Theta_{ab}$  of the outgoing electrons for
incident electron energy $E_i=15.6$ eV vs. ejection angle
$\theta_a$ of the slow outgoing electron. Continuous curves,
present results ; dashed-curves, ECS results [9, 50]; dash-dotted
curves, CCC results [26, 48]. The experimental results are the
absolute measured values of R\"oder \textit{et al} [47, 48].
Present results have been multiplied by a factor 0.5
(see text).\\
\noindent
    \textbf{Figure 1(b).} TDCS in coplanar equal-energy-sharing
geometry for incident electron energy $E_i=15.6$ eV  for fixed
$\theta_a$ and variable $\theta_b$ of the out going electrons.
Continuous curves, present results ; dashed-curves, ECS results
[9, 50]; dash-dotted curves, CCC results [26, 48]. The
experimental results are the absolute measured values of R\"oder
\textit{et al} [47,48]. Present results have been multiplied by a
factor 0.5
(see text).\\
\noindent
     \textbf{Figure 1(c).} TDCS in coplanar equal-energy-sharing
with two electrons emerging on opposite sides of the direction of
the incident electron with equal angle $\theta_a$ and energy
$E_i=15.6$ eV . Continuous curves, present results ;
dashed-curves, ECS results [9, 50]; dash-dotted curves, CCC
results [26,48]. The experimental results are the absolute
measured values of R\"oder \textit{et al} [47, 48]. Present
results have been multiplied by a factor 0.5 (see text).\\

\noindent
    \textbf{Figure 2(a).} Same as in figure 1(a) but for
17.6 eV incident electron energy. The experimental results are the
recent absolute measured values of R\"oder \textit {et al}[51] and
the CCC results are as in [29]. Here the present results are free
from any multiplicative factor.\\
\noindent
    \textbf{Figure 2(b).} Same as in figure 1(b) but for
17.6 eV incident electron energy. The experimental results are the
recent absolute measured values of R\"oder \textit {et al}[51] and
the CCC results are as in [29] . Here the present results are free
from any multiplicative factor.\\
\noindent
    \textbf{Figure 2(c).} Same as in figure 1(c) but for
17.6 eV incident electron energy. The experimental results are the
recent absolute measured values of R\"oder \textit {et al}[51] and
the CCC results are as in [29]. Here the present results are free
from any multiplicative factor.\\

\noindent
    \textbf{Figure 3.} Same as in figure 2(a) but for 19.6 eV
incident electron energy. The relative measured results of R\"oder
\textit {et al} [47-49] are normalized as in [9].\\

\noindent
    \textbf{Figure 4.} Same as in figure 3 but for 25 eV
incident electron energy. The CCC results are from [28, 49] \\

\noindent
    \textbf{Figure 5.} Same as in figure 3 but for 30 eV
incident electron energy. \\

\end{document}